# Low Noise Microwaves for Testing Fundamental Physics

Eugene N. Ivanov and Michael E. Tobar

*Department of Physics, University of Western Australia, 35 Stirling Hwy., Crawley, 6009*We studied noise properties of microwave signals transmitted through the cryogenic resonator. The experiments were performed with the 11.342 GHz sapphire loaded cavity resonator cooled to 6.2 K. Based on the measured transmission coefficient of the cryogenic resonator we computed its noise suppression function. This was done via Monte-Carlo simulations some details of which are discussed in this Letter. Next, we measured technical fluctuations of a signal incident on the cryogenic resonator. Having processed these data with the previously computed noise filtering "template" we inferred noise spectra of the transmitted signal. We found that spectral densities of both phase and amplitude fluctuations of the transmitted signal were close to the thermal noise limit of -180 dB/Hz at Fourier frequencies F≥ 10 kHz. Such thermal noise limited microwaves allow more precise tests of special relativity and could be useful at some stages of quantum signal processing.Currently, microwave signals with the best long-term frequency stability are derived from the optical sources[1]. This is the result of the outstanding progress in laser frequency stabilisation[2-4] and optical frequency synthesis[5,6]. Not only the long-term frequency stability of the optical clock is preserved when its signal is transferred to the microwave domain; fast phase fluctuations of the optical signal are also strongly reduced[7-10]. For example, X-band microwave signals with Single-Sideband (SSB) phase noise of -173 dBc/Hz at Fourier frequencies near 10 kHz were synthesized using femtosecond laser technology[9]. This noise performance is comparable to what was achieved with interferometric Sapphire Loaded Cavity (SLC) oscillator in the same range of Fourier frequencies[11].

In this work we show that, a classical loop oscillator based on a cryogenic SLC resonator offers better noise performance than the current state-of-the-art at Fourier frequencies above 1 kHz. We also show that noise spectra of the cryogenic oscillators are essentially free from vibration-induced phase/amplitude disturbances inherent to the photonic systems[9]. The cryogenic sapphire oscillators are well-suited to precision tests of fundamental physics, qubit control[12] and laboratory search for Dark Matter[13-15].

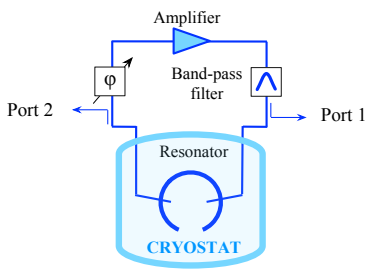

Fig. 1. Schematic diagram of cryogenic loop oscillator

Fig. 1 shows schematic diagram of the microwave loop oscillator. Its key components are an amplifier and a resonator. If the latter is a multi-mode resonant system (such as the SLC resonator) a band-pass filter is needed to eliminate the mode competition. A variable phase-shifter φ in Fig. 1 enables tuning of oscillator frequency $f_{osc}$ relative to the SLC resonant frequency $f_{res}$. The range of oscillator frequency tuning is limited by the loop gain margin and is typically of the order of the resonator linewidth.

At $f_{osc} = f_{res}$, Power Spectral Density (PSD) of phase fluctuations at the output of loop amplifier (port 1 in Fig. 1) is given by[16]

$$S_\phi^{(1)}(\mathcal{F}) = S_f^{res}(\mathcal{F}) \frac{1}{\mathcal{F}^2} + S_\phi^{amp}(\mathcal{F})\left[1 + \left(\frac{\Delta f_{0.5}}{\mathcal{F}}\right)^2\right] \quad (1)$$

where $\Delta f_{0.5}$ in is the half linewidth of the resonator, $S_\phi^{amp}$ is the PSD of amplifier phase fluctuations and $S_f^{res}$ is the PSD of resonator frequency fluctuations (induced by fluctuations of resonator temperature and dissipated microwave power).

Eq. 1 was derived assuming the resonator with Lorentzian transmission coefficient. In such a case, spectral density of phase fluctuations of the transmitted signal (at port 2 in Fig. 1) is expressed as the following sum

$$S_\phi^{(2)}(\mathcal{F}) \simeq S_\phi^{(1)}(\mathcal{F}) \frac{1}{1+(\mathcal{F}/\Delta f_{0.5})^2} + \frac{k_B T_o}{P_{trans}} \quad (2)$$

where $k_B$ is the Boltzmann constant, $T_o$ is the ambient temperature and $P_{trans}$ is the power of the transmitted signal. In the above equation, the first term is the spectral density of the low-pass filtered phase fluctuations of the incident signal; the second term is the spectral density of thermal phase noise.

A simple relation between two phase noise spectra given by the first term in Eq. 2 does not hold for the real microwave resonators whose transmission profiles are rarely Lorentzian. This is especially true for the cryogenic SLC resonators where the high-Q modes often exist as doublets[17,18]. Thus, Fig. 2 shows amplitude and phase transfer functions of the cryogenic SLC resonator used in our experiments. Its resonant frequency and 3dB linewidth are, respectively, 11.343 GHz and 58 Hz. The shape of the amplitude transfer function is close to the Lorentzian only within +/- 0.5 kHz around the resonance. In such a case, we show that the relationship between spectral densities of the incident and transmitted signals can be found from numerical simulations, as discussed below.

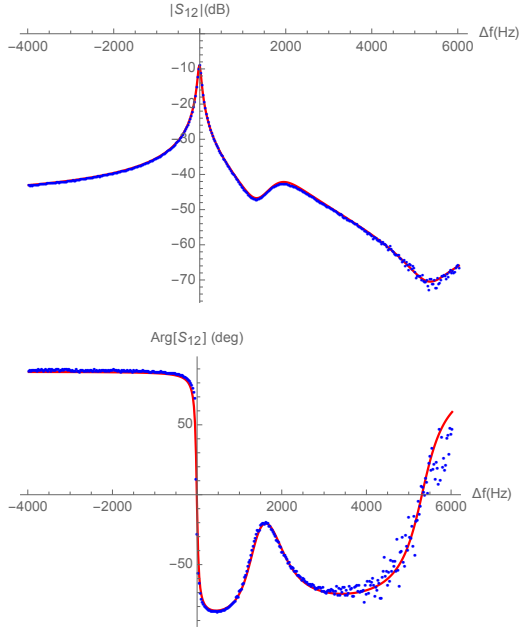

Fig. 2. Magnitude (top) and phase (bottom) of complex transmission coefficient of the cryogenic resonator vs offset from 11.343 GHz. Red trace is a fit obtained from the lumped element model of the cryogenic resonator.

A flowchart of an algorithm for calculating phase fluctuations of a signal transmitted through a resonator is shown in Fig. 3. Here, the transmitted signal is mixed with two replicas of the incident signal phase-shifted by π/2.

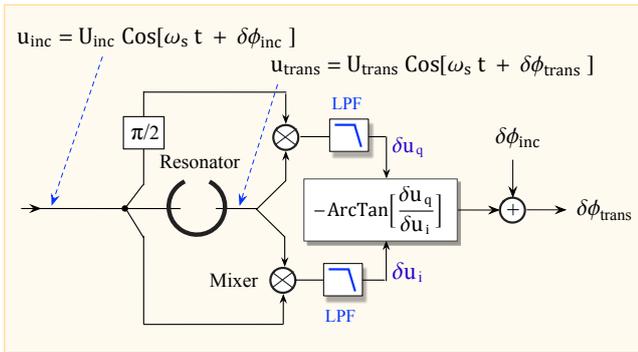

Fig. 3. Homodyne detection of phase fluctuations of the transmitted signal

The mixing products (beat notes) are low-pass filtered giving rise to slowly-varying random processes $\delta u_i$ and $\delta u_q$. The ratio of those processes determines additional phase fluctuations acquired by incident signal upon transmission through the resonator: $\delta\phi_{add} = -\text{ArcTan}[\delta u_q/\delta u_i]$.
Denoting phase fluctuations of the incident signal as $\delta\phi_{inc}$, for the transmitted signal we can write:

$$\delta\phi_{trans} = \delta\phi_{inc} - \text{ArcTan}[\delta u_q/\delta u_i] \qquad (3)$$

Since the $\delta\phi_{inc}$, is the "user-defined" random process, we can compute both $\delta\phi_{trans}$ and resonator Noise Suppression Factor (NSF) defined as a ratio of two spectral densities

$$\text{NSF} = S_\phi^{trans}(\mathcal{F}) \big/ S_\phi^{inc}(\mathcal{F}) \qquad (4)$$

Having introduced the NSF, we can rewrite Eq. 2 as

$$S_\phi^{(2)}(\mathcal{F}) \simeq S_\phi^{(1)}(\mathcal{F})\,\text{NSF} + k_B T_o / P_{trans} \qquad (5)$$

Fig. 4 shows the NSF as a function of Fourier frequency computed for given cryogenic SLC resonator. A step-like distortion at the NSF plot is due to the "hump" at the resonator amplitude transmission a few kHz away from the resonance peak (Fig. 2). At F = 1 kHz, the NSF is close to 30 dB. This is not very different from the estimate obtained from Eq. 2 for the Lorentzian resonator with the same linewidth as that of the cryogenic resonator used.

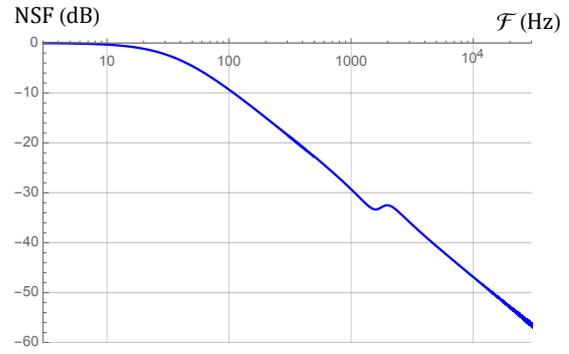

Fig. 4. Noise suppression factor vs Fourier frequency

In our calculations, we relied on the Discrete Fourier Transforms (DFT) algorithms implemented in Wolfram's Mathematica. Simulations were carried out in the Time Domain with phase fluctuations of the incident signal modelled as a Gaussian process with "white" power spectral density. Magnitude of phase fluctuations was assumed to be small ($\delta\phi_m \ll 1$) to avoid saturation of the phase detectors (mixers) in Fig. 3. Duration of the incident signal (i.e. observation time) was chosen from the condition: $T_{obs} \sim 10/\Delta f_{0.5} \sim 0.3$ s to allow at least twenty DFT points within the linewidth of the cryogenic resonator. Such choice of signal duration minimized errors associated with the DFT picket-fence effect[19].

The key parameter of simulations is the sampling frequency $f_{sample}$. If chosen formally ($f_{sample} \gg f_{res}$) it would result in prohibitively long computation times with the number of samples ($N_{samples} = T_{obs} f_{sample}$) running into tens of billions. Fortunately, the nature of phase fluctuations in the cryogenic loop oscillator, in particular the rapid decrease of noise spectral density with Fourier frequency, eliminates the need for the high sampling rates. This can be understood by examining the phase noise spectra of the cryogenic sapphire oscillator (see Fig. 5). Here, the top trace corresponds to the phase noise spectrum of the incident signal. It was measured using a classical two-oscillator technique[20]. For that, a second cryogenic sapphire oscillator with noise properties similar to the first one was constructed. Frequency separation between two oscillators was approximately 39 MHz. Phase fluctuations of the 39 MHz beat note were characterized with the 5125A phase noise test set (originally from Symmetricom). The phase noise data in Fig. 5 are presented as the SSB power spectral densities in dB/Hz. Phase noise spectra of the individual oscillators are shown.

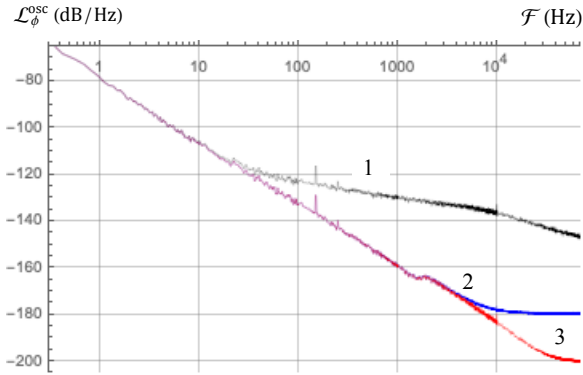

Fig. 5. SSB PM-noise spectra of the 11.343 GHz cryogenic loop oscillator. Trace 1 corresponds to signal incident on resonator. Traces 2 and 3 correspond to the transmitted signals at T = 296 K and 6.2 K respectively.

The SSB phase noise of the incident signal is close to -140 dB/Hz at F = 10 kHz. This is approximately 40 dB higher than the thermal noise limit for a signal with power $P_{trans}\sim 3$ dBm. On the other hand, the noise suppression factor of Lorentzian resonator with linewidth $2\Delta f_{0.5} \sim 58$ Hz is nearly 50 dB at the same Fourier frequency of 10 kHz (Eq. 2). This means that noise filtering action of the cryogenic resonator is sufficiently strong to reduce phase fluctuations of transmitted signal to the thermal noise floor at F ≥10 kHz. Avoiding the need to compute the NSF at high Fourier frequencies is one reason why numerical simulations can proceed at relatively low sampling rates. Another reason to further reduce the sampling rate is due to the fact that the NSF depends on the shape of resonator transmission peak; it is first order independent on peak's location along the frequency axis. For example, the NSF of the Lorentzian resonator depends only on its linewidth and Fourier frequency.

Overall, in our analysis we assumed: $F_{max}$ = 30 kHz and $f_{res}$ = 30 MHz. As for the sampling frequency, it was chosen from $f_{sample}$ = 16 $f_{res}$ to minimize the DFT leakage errors[19].

Trace 2 in Fig. 5 shows inferred phase noise spectrum of the transmitted signal. The thermal noise floor of -180 dB/Hz is reached at F~10 kHz. This noise floor corresponds to $P_{trans}$= 2 mW, which is the power measured at the output of the directional coupler attached to the room temperature interface of the cryostat. Power of microwave signal entering the cryostat was 80 mW. Most of it was lost in the transmission lines leading to and from the SLC resonator. Approximately 20 mW were dissipated in sapphire crystal raising its temperature to ~6.2 K. Power of the transmitted signal exiting the cryostat was slightly less than 10 mW.

The NSF is an inherent property of the resonator. The same NSF used for inferring spectral density of phase fluctuations of the transmitted signal can be applied to evaluation of its amplitude noise. Fig. 6 shows SSB spectra of amplitude fluctuations of the cryogenic SLC oscillator. Trace 1 in Fig. 6 corresponds to AM-noise spectrum of the incident signal.

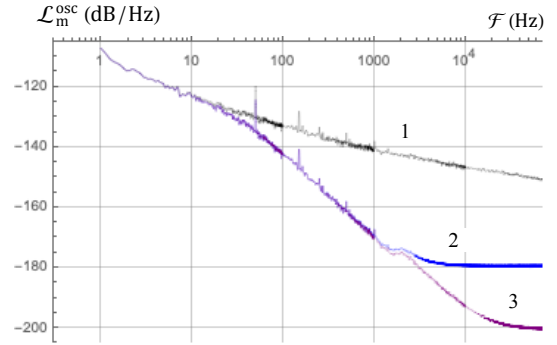

Fig. 6. SSB AM-noise spectra of the 11.343 GHz cryogenic loop oscillator. Trace 1 corresponds to signal incident on resonator. Traces 2 and 3 correspond to transmitted signal at T = 296 K and 6.2 K respectively.

Traces 2 and 3 give AM-noise spectra of the transmitted signals at ambient temperatures of 296 K and 6.2 K, respectively. The latter case corresponds to the situation where the signal transmitted does not have to leave the cryogenic environment.

The noise spectra in Figs. 5 and 6 show that cryogenic oscillators are capable of generating microwave signals with very low levels of technical fluctuations. Still, there are ways to further improve spectral purity of microwave signals. Thus, a simple replacement of the FET amplifiers used in our experiments with their BJT counterparts should reduce oscillator phase noise by additional 5…7 dB. Significantly lower levels of phase/amplitude fluctuations can be reached with implementation of the Reduced Noise Amplifiers (RNA)[21]. We are currently working on characterisation of noise properties of the RNA and on its integration into the loop oscillator. The results of our preliminary measurements indicate that the SSB phase noise of the RNA-based loop oscillator would be close to the thermal noise limit of -180 dBc/Hz at 300 Hz instead of the current value of 10 kHz[22]. The noise properties of the cryogenic loop oscillator based on the RNA will be presented in our next submission.

We acknowledge useful discussions with NIST, Boulder Labs researchers J. Bergquist, D. Hume, C. Oates, S. Diddams and F. Quinlan. We also acknowledge financial support of the Australian Research Council (grants DP190100071 and CE170100009)

**DATA AVAILABILITY**
The data that support the findings of this study are available from the corresponding author upon reasonable request.